# Laser writing of individual atomic defects in a crystal with near-unity yield.

Yu-Chen Chen[1†], Benjamin Griffiths[1,2], Laiyi Weng[1], Shannon Nicley[1], Shazeaa N. Ishmael[1], Yashna Lekhai[3], Sam Johnson[1], Colin J. Stephen[3], Ben L. Green[3], Gavin W. Morley[3], Mark E. Newton[3], Martin J. Booth[2], Patrick S. Salter[2], and Jason M. Smith[1*]

[1] Department of Materials, University of Oxford, Parks Road, Oxford OX1 3PH, UK

[2] Department of Engineering Science, University of Oxford, Parks Road, Oxford OX1 3PJ, UK

[3] Department of Physics, University of Warwick, Coventry CV4 7AL, UK

*Correspondence to: jason.smith@materials.ox.ac.uk.

† Current address – 3 Physikalisches Institut, Universität Stuttgart, 70550 Stuttgart, Germany

**Abstract**: Atomic defects in wide band gap materials show great promise for development of a new generation of quantum information technologies, but have been hampered by the inability to produce and engineer the defects in a controlled way. The nitrogen-vacancy (NV) color center in diamond is one of the foremost candidates, with single defects allowing optical addressing of electron spin and nuclear spin degrees of freedom with potential for applications in advanced sensing and computing. Here we demonstrate a method for the deterministic writing of individual NV centers at selected locations with high positioning accuracy using laser processing with online fluorescence feedback. This method provides a new tool for the fabrication of engineered materials and devices for quantum technologies and offers insight into the diffusion dynamics of point defects in solids.

**Main Text:**

The engineering of materials at the scale of individual atoms has long been viewed as a holy grail of technology. With the extreme miniaturization of modern semiconductor technology to sub- 10 nm feature sizes and the emerging promise of quantum technologies that rely inherently on the principles of quantum physics, the ability to fabricate and manipulate atomic-scale systems is becoming increasingly important.

One promising approach to quantum technologies is the use of 'color center' point defects in wide band gap materials that display strong optical transitions, allowing the addressing of single atoms using optical wavelengths within the transparency window of the solid. Fluorescence from single color centers displays quantum statistics with potential for use in communications, sensing and metrology (*1*), while some color centers also possess spin degrees of freedom that can be accessed via the optical transitions, opening the door for use as highly sensitive magnetic field sensors (*2*) or as quantum memories for use in optical networks for communications or computing technologies (*3, 4*).

The fabrication and engineering of color centers is challenging, since they generally comprise compound defects containing one or more 'elements' – impurity atoms and lattice vacancies – which are bound together in a stable configuration. To build them atom-by-atom would be painfully slow (if even possible), so the process of fabrication involves the introduction of the required elemental defects into the lattice followed by thermal annealing to stimulate diffusion of one or more element through the lattice whereupon random binding occurs to create the color center.

The most extensively researched color center is the negatively charged nitrogen-vacancy (NV) center in diamond, for which measurements on single defects were first reported by researchers at the University of Stuttgart in 1997 (*5*). NV centers have since formed a basis for advances in photon-mediated entanglement (*6*), quantum teleportation over long distances (*7*), and nanoscale nuclear magnetic resonance (*8*), and are one of the few physical systems to have been shown to support quantum logic gates with fidelity above the threshold for fault-tolerant quantum computing (*9*). NV centers are formed in diamond by the binding of a substitutional nitrogen impurity with a lattice vacancy along a [111] crystal axis. The controlled positioning of NV centers therefore requires the targeted implantation of either nitrogen ions or vacancies in the crystal, which are typically achieved using ion implantation and electron beam irradiation methods (*10-14*). Recently, laser processing was shown to be an effective way of writing lattice vacancies into diamond with minimal residual damage such that binding with substitutional nitrogen impurities already present in the lattice yielded high quality single NV centers (*15,16*). The 'global' thermal anneal used to facilitate the random diffusion of vacancies and subsequent NV center creation meant that the number of NV centers produced per site was determined by Poisson statistics such that the maximum probability of creating a single NV center at a given site was 37%, and that the positioning accuracy of the resultant NV centers was limited to a few hundred nanometers by the vacancy diffusion length. This low yield and modest positioning accuracy would potentially limit the usefulness of the technique in device manufacturing.

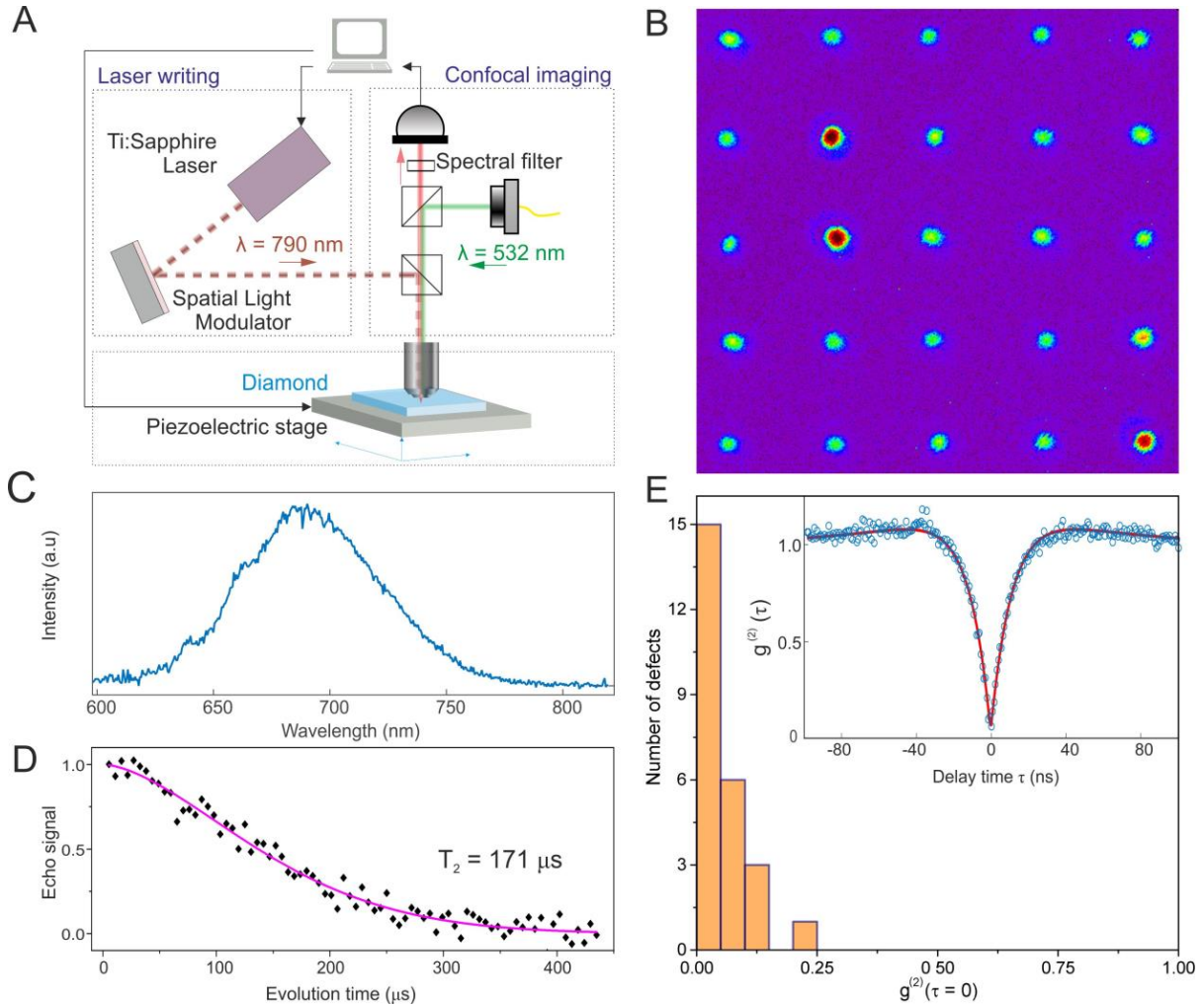

**Fig. 1. Deterministic laser writing of NV color centers in diamond.** (**A**) Schematic of the laser writing apparatus. (**B**) Fluorescence image of a high-yield 5 x 5 array of single NV centers on a 2 μm square grid. (**C**) Typical fluorescence spectrum and (**D**) measured spin echo data from a written color center (see Methods for decay function). (**E**) Histogram of $g^{(2)}(0)$ values for the array in (B), (inset) a typical $g^{(2)}(\tau)$ dataset corrected for background signal (see Methods).

In this report we enter a distinct regime in the fabrication of these atomic-scale defects. We use laser processing to achieve near-deterministic writing of NV centers in bulk diamond with an accuracy of around 50 nm in the image plane. The enabling capabilities are two-fold. Firstly the use of laser processing to not only create but also to diffuse vacancies in place of a thermal anneal, providing site-specific control of NV center formation; and secondly the use of online fluorescence measurement that provides feedback to allow active control of the process (Figure 1A, for detailed optical layout see Figure S1).

Figure 1B shows a fluorescence image of an array of 25 sites processed in a single crystal diamond sample (see Methods for sample details). At each site in the array, an initial laser pulse of energy 27 nJ (the 'seed pulse') was used to generate vacancies, followed by a 1 kHz stream of pulses of energy 19 nJ ('diffusion pulses') to induce vacancy diffusion. Fluorescence within the wavelength range 650 nm to 750 nm was monitored until a signal indicating the creation of a negatively charged NV center was recorded, at which point the processing was halted.

Measurements of fluorescence spectra (Fig 1C), optically detected electron spin echo (Fig 1D), fluorescence polarisation (Figure 4D and Supplementary Information Figure S3) and photon statistics (Fig 1E and Figure S4) from the processed sites reveal that 24 of the 25 sites contain a single NV center, a yield of 96%, with just one site of the 25 (4$^{th}$ row from top, right-hand column) containing two NV centers. The NV centers are stable in the negatively charged state with no evidence of fluorescence from the charge-neutral state NV$^0$. They display distinct photon anti-bunching (see Methods), with $g^{(2)}(0) < 0.2$ after correction for background fluorescence and no discernible fluorescence from other defects such as the isolated vacancy (GR1) or extended defects (B band). Electron spin coherence ($T_2$) times of up to 170 µs were recorded using Hahn echo measurements (see Methods).

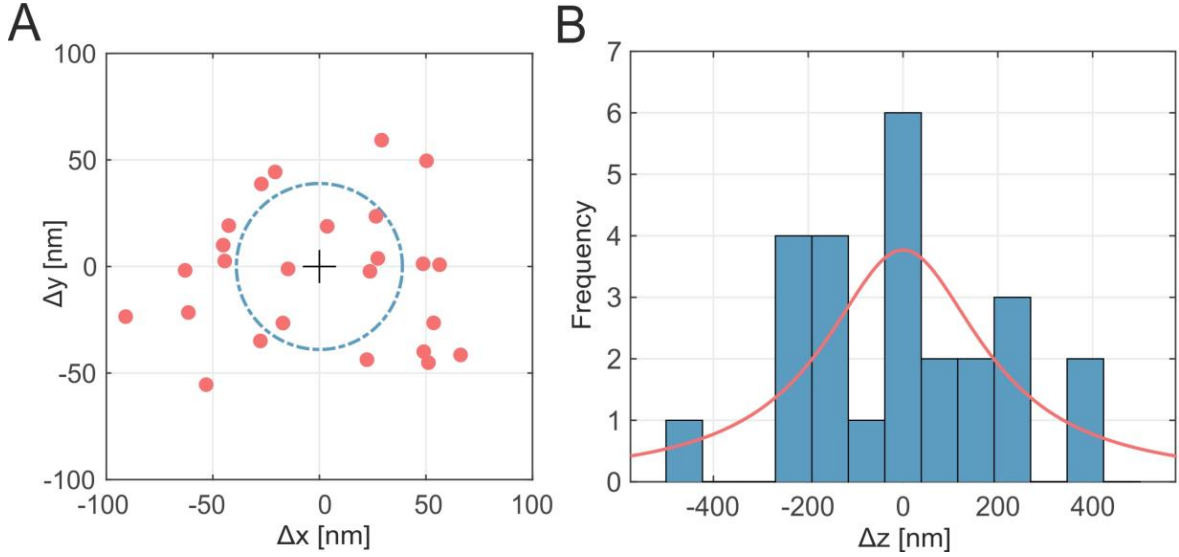

**Fig. 2. Positioning accuracy of single NV centres. (A)** Scatter plot of the fitted positions of the NV centers in Figure 1B relative to the target grid in the image plane. The dashed circle represents the deviation parameter σ for the best fit 2D Gaussian distribution function. **(B)** Histogram of the fitted depths of the NV centers (see Methods) with a best-fit Lorentzian distribution.

The positions of the NVs in the image plane relative to the targeted array points were measured with an accuracy of 20 nm by establishing the centroids of the points in the fluorescence image (see Methods). Figure 2A shows the spatial distribution obtained. A maximum likelihood analysis using the 2D Gaussian probability distribution $P(\Delta x, \Delta y, \sigma) = \frac{1}{2\pi\sigma^2} e^{-(\Delta x^2 + \Delta y^2)/2\sigma^2}$ reveals a deviation parameter of $\sigma = 39 \pm 8$ nm, indicated by the dashed circle in the figure. This measured scatter in position relative to the targeted points is a factor of 3.7 smaller than the equivalent width of the focal spot of the writing laser (see Methods), and a factor of 3.6 improvement over the in-plane positioning accuracy achieved in (*15*). The relative depth of each NV center below the sample surface was measured with an accuracy of about 50 nm from sectional fluorescence images recorded with the confocal microscope. Figure 2B shows the measured distribution, together with a fitted Lorentzian line shape of half-width-at-half-maximum 203 ± 74 nm. This depth variation is a factor of 4.2 smaller than the estimated Rayleigh range of of the focal spot of the writing laser (see Methods).

An exemplar trace from the fluorescence intensity monitor shown in Figure 3 provide insight into the dynamics of NV center generation. The zero of the horizontal time axis corresponds to the time of the seed pulse, labelled (i), after which the lower energy 1 kHz pulses commence.

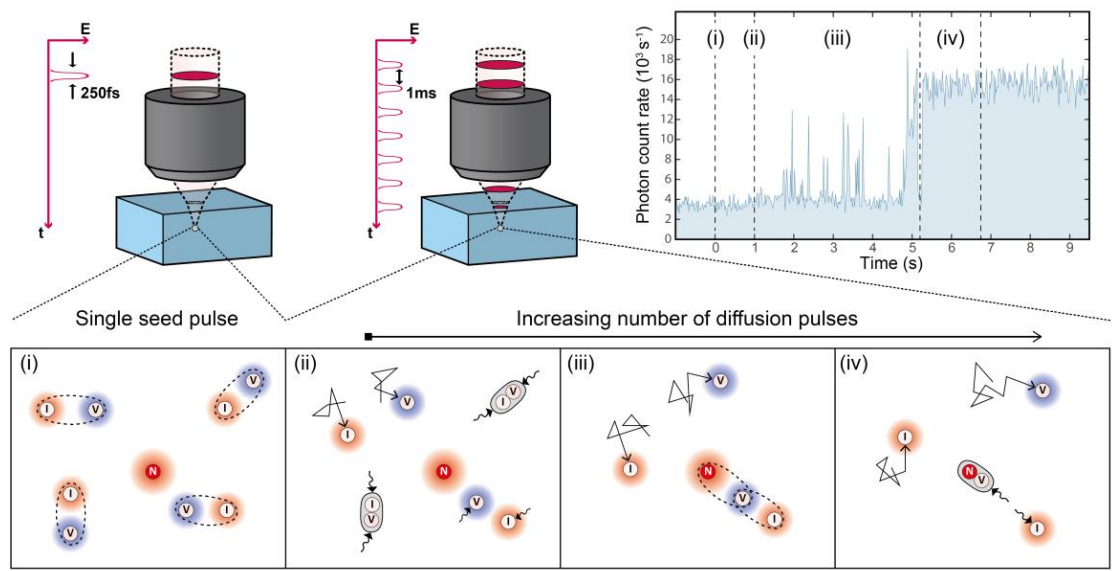

**Fig. 3. Diffusion dynamics of point defects in NV center creation.** The proposed mechanism for creation of stable NV centers occurs in four stages. (i) Intense seed pulse from laser generates Frenkel defects; (ii and iii) lower energy pulses at 1 kHz cause diffusion and recombination of Frenkel defects – some Frenkel defects align with substitutional nitrogen impurities whereby vacancy binds occasionally to nitrogen and intermittent NV fluorescence is observed; (iv) interstitial diffuses away from NV pair and stable fluorescence is observed.

The fluorescence intensity is recorded in photon detection events per second, with an integration time of 20 ms per measurement. Commencement of the 1 kHz pulse train at (ii) is closely followed by intermittent fluorescence of a higher intensity in region (iii), before stable, high intensity fluorescence is observed (region (iv)) and the process terminated.

This intermittent signal is interpreted as unstable binding of the vacancy to the substitutional nitrogen in the presence of a nearby carbon split-interstitial defect. Panels i-iv in Figure 3 correspond to the regions of the monitor trace identified above, and depict the proposed NV formation process. The vacancy and split interstitial are the component parts of Frenkel defects formed by the initial seed pulse (panel i). These two defects have relatively low activation energies for diffusion, of 2.3 eV and 1.6 eV respectively (*17-19*), and are both mobile under the 19 nJ diffusion pulses, while the substitutional nitrogen atoms, with activation energy of 5 eV (*20*), are highly immobile and can be considered to remain at fixed lattice points. The diffusive motion of the vacancy and interstitial is affected by the presence of strain fields surrounding the individual point defects. The vacancy is a point of lattice depletion so it is surrounded by a region of tensile (negative) strain while the split-interstitial and substitutional nitrogen add bulk to the lattice and so are surrounded by regions of compressive (positive) strain. Strain fields behave like electrostatic fields due to charges – same-sign fields correspond to repelling forces between the defects and opposite-sign fields correspond to attractive forces – such that the vacancy is attracted to both the substitutional nitrogen and to the interstitial, while the interstitial is repelled by the nitrogen. Frenkel defects created far from a nitrogen may either recombine or dissociate (panel ii) while for those created near to a nitrogen the vacancy will tend to locate between the interstitial and the nitrogen, whereby its diffusive motion results in occasional but unstable binding to the nitrogen, and the observed intermittent NV fluorescence (panel iii). The NV complex does not provide a strong attractive force to the interstitial, so that the interstitial may diffuse far enough away that the strain fields no longer interact, whereupon the vacancy remains bound to the nitrogen and stable NV fluorescence is observed (panel iv).

Continued processing of a site beyond the creation of a single NV center can reveal rich dynamics offering additional potential for device engineering. Figure 4A shows an example of a fluorescence trace in which a second NV center is formed, while Figure 4B shows a monitor trace in which an NV center is created, then destroyed, before another defect is created at a later time. Such processes allow for a degree of selection of NV properties. The simplest example is that of NV orientation in the lattice. The orientation of each NV center can be ascertained by the optical polarization of its fluorescence, since the transition dipoles of the defect lie in the plane perpendicular to the physical axis of threefold rotational symmetry. In the sample used the crystal plane parallel to the surface is (110) such that NVs whose axes of symmetry are oriented along the $[1\bar{1}1]$ and $[\bar{1}11]$ crystal axes lie in the plane and their fluorescence displays a high degree of linear polarization. NVs oriented oriented along $[111]$ and $[11\bar{1}]$, by contrast, make angles of 55° with the image plane, so that their fluorescence is stronger but only weakly polarized (Figure 4C and D, and S3). By terminating processing when fluorescence was observed within a specific window of intensity, it was therefore possible to exercise some control over the statistical distribution of orientations in the lattice. The array in figure 1 was fabricated with no selection of fluorescence intensity and 23 of the 26 NV centers (88%) created were in one of the two orientations with their axes lying in the plane of the sample (Figure S3), only three (12%) were created in the two orientations with their axes out-of-plane (the reason for this statistically significant bias is presently unknown). By terminating processing only when a fluorescence signal intensity in the range 11-15 kc/s was observed, the fraction of single NV centers lying with axes out-of-plane increased to 87% (Figure 4E). A third array in which processing was terminated when a signal of 8-11 kc/s was recorded revealed 84% of NV centers in the in-plane orientations.

We now discuss briefly the physical mechanism behind the laser processing method used. The generation of vacancies in diamond by a sub-picosecond laser pulse is thought to be via multi-photon absorption that promotes electrons to high energy states, followed by rapid relaxation of the electrons by transfer of their energy to the lattice (*21*). It follows that the laser-induced vacancy diffusion process observed here may occur by a similar process, but with a lower nonlinearity resulting from the lower energy requirement for a vacancy to hop to an adjacent lattice site than for a new vacancy to be created. Our experiments suggest a nonlinearity as high as 15 for generation of Frenkel defects and of up to 10 for vacancy diffusion, such that only a narrow window of diffusion pulse energies exists (13-19 nJ) in which NV centers are generated but continued processing does not ultimately lead to runaway lattice damage. These observations in the bulk of the diamond contrast with recent work by Kononenko *et al.* which revealed a logarithmic dependence of NV generation over a wide range of laser fluence during nanoablation of the diamond surface (*22*).

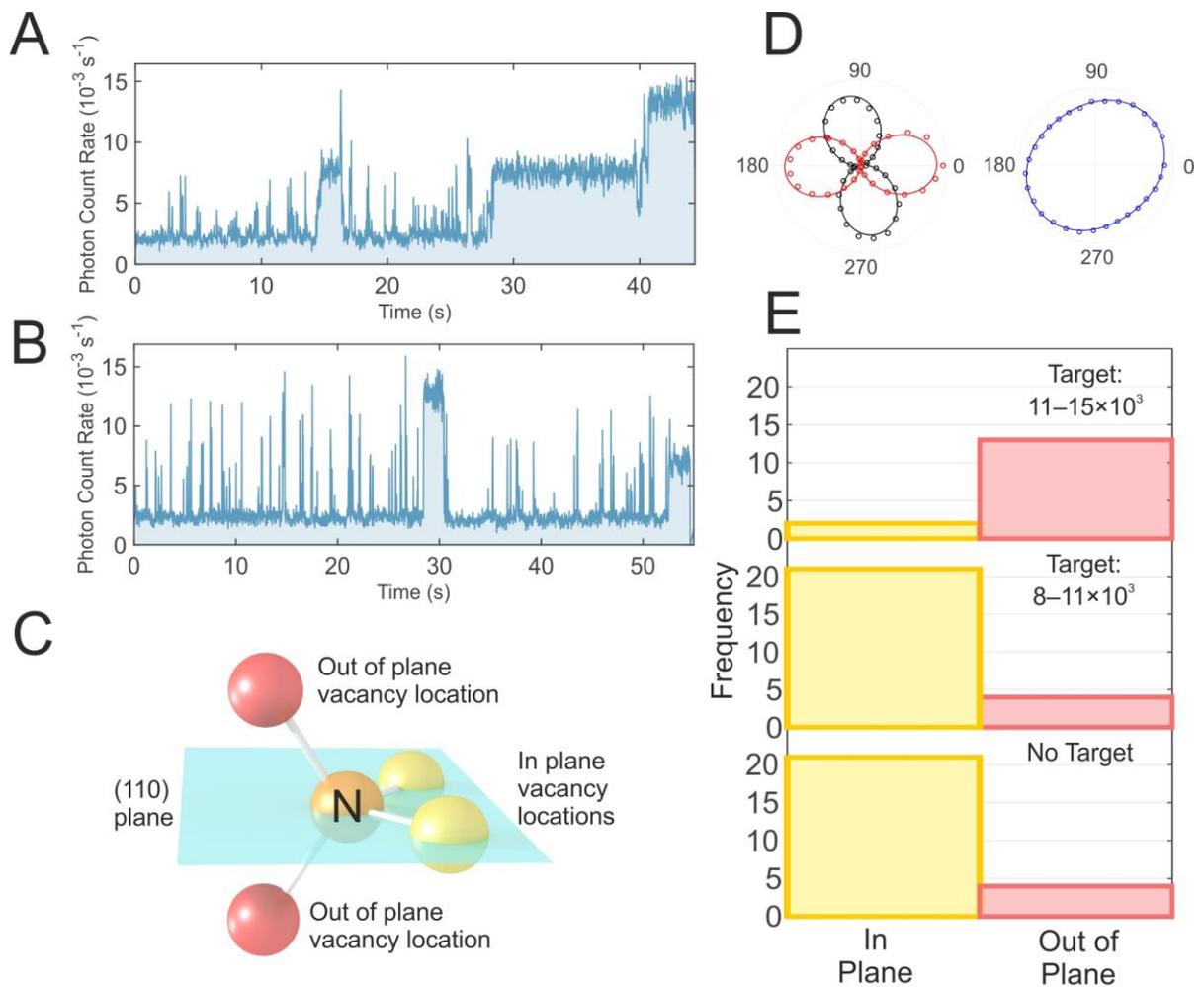

**Figure 4: Advanced control of NV properties.** (**A**) Fluorescence monitor trace showing creation of two NV centers at a processing site. (**B**) Trace showing the creation and destruction of an NV center (time = 28-30 seconds) followed by selection of the second defect created. (**C**) Schematic of the four orientations of NV centers relative to the (110) image plane. (D) Example polarization datasets from in-plane and out-of-plane NV centers. (E) Histograms of orientations of defects in array under different termination conditions.

In summary we have reported a method for localized laser processing of wide band gap materials and the deterministic creation of single NV centers at desired locations in diamond. Feedback provided by online fluorescence monitoring provides information on defect formation, and can be used to identify a range of other properties of the color centers created, providing additional routes to device engineering. The processing method can in principle be applied to the controlled writing of other color centers in diamond, or fluorescent defects in other materials such as the di-vacancy in silicon carbide (*23*). It provides a tool with potential for wide application in the engineering of materials at the quantum level.

**Acknowledgments:**

We would like to acknowledge DeBeers and Element Six for providing suitably characterized diamond samples for this work, and in particular Daniel Twitchen and David Fisher for their comments on the manuscript. Data reported in the paper are presented in the Supplementary Materials and are archived at (tbc). The work was funded by the UK Engineering and Physical Sciences Research Council (EPSRC) through the UK hub in Networked Quantum Information Technologies (NQIT), grant # EP/M013243/1.

Y-CC, B Griffiths and SN carried out the experiments and performed the data analysis with supervision from JS and PS; B Griffiths, LW, SJ and PS constructed the laser writing and fluorescence feedback apparatus; SI, YL, CJS and BLG carried out the Hahn echo and spatial localization measurements under the supervision of GM and MN; and JS, YCC, PS and MB conceived the experiment; all authors contributed to writing the manuscript.

## Supplementary Materials:

### Materials and Methods:

The samples used were single-crystal type 1b diamond with nitrogen concentration of 1.8 ppm, produced by a High Pressure High Temperature (HPHT) technique. The diamond was cut and polished with flat surfaces parallel to the (110) plane of the cubic crystal.

The optical layout for the combined laser processing and fluorescence feedback apparatus is shown in Figure S1. The laser processing was performed using a regeneratively amplified Ti:Sapphire laser (Spectra Physics Solstice) at a wavelength of 790 nm and a 1 kHz pulse repetition rate. The laser beam was expanded onto a liquid crystal phase-only spatial light modulator (SLM) (Hamamatsu X10468-02), which was imaged in a 4f configuration onto the back aperture of a 60× 1.4NA Olympus PlanApo oil immersion objective. The SLM provided the capability to perform aberration correction to compensate for the effects of refraction at the surface of the diamond. The diamond sample was mounted on precision translation stages (Aerotech x-y: ABL10100; z: ANT95-3-V) providing three dimensional control. Prior to the objective the laser pulse was linearly polarized and had a duration which was measured to be 250 fs using an intensity autocorrelator (APE Pulsecheck). The pulse duration at focus will be slightly increased due to dispersion in the objective lens. To optimize the aberration correction, the phase pattern displayed on the SLM was adjusted to minimize the pulse energy needed to produce visible fluorescence at test processing positions of similar depth in the sample.

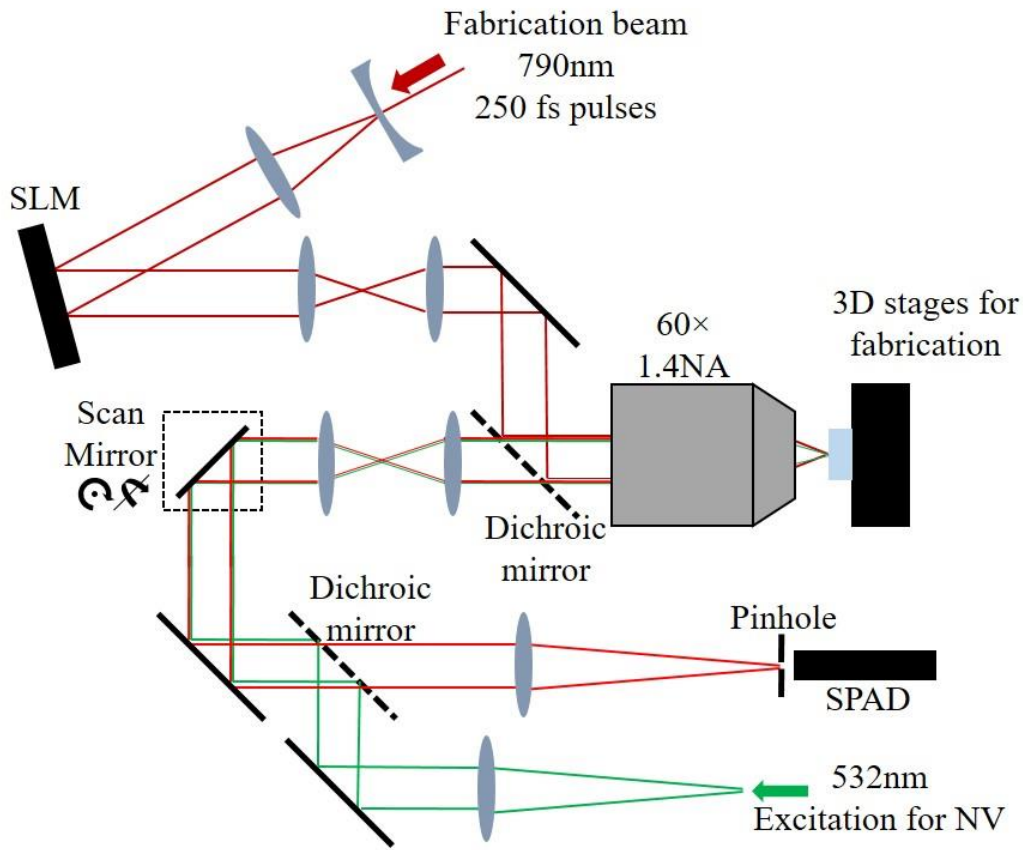

Figure S1: Optical layout of the laser writing and fluorescence feedback apparatus. Telecentric lens configurations are used to image the spatial light modulator (SLM) and scanning mirror onto the back aperture of the objective lens.

Fluorescence feedback was achieved using a custom built scanning confocal photoluminescence (PL) microscope, with excitation from a frequency-doubled diode-pumped YAG laser at 532 nm and detection using a silicon single photon avalanche diode (Excelitas). Excitation power of 2 mW was reflected towards the sample at a 570 nm dichroic beamsplitter, and fluorescence transmitted through the dichroic was additionally filtered using a 650-750 nm band pass filter.

The point spread function of the processing laser is that of a focused Gaussian beam

$$I(r,z) = I_0 \cdot \frac{1}{1 + \left(\frac{z}{z_R}\right)^2} \cdot e^{-\frac{2r^2}{w_z^2}}$$

where $w_z = w_0 \sqrt{1 + \left(\frac{z}{z_R}\right)^2}$ is the beam width at axial displacement $z$, $w_0$ is the beam waist and $z_R$ is the Rayleigh range. Based on the numerical aperture of the objective lens we estimate that $w_0 = 297$ nm and $z_R = 852$ nm.

Fluorescence imaging was carried out using a confocal PL microscope and spectroscopy with a 500 mm spectrograph (Acton SpectraPro 500i) fitted with a back-illuminated CCD camera (Princeton Spec-10 100B), with > 90% quantum efficiency across the wavelength range of interest. Excitation was performed using a frequency-double YAG laser ($\lambda$=532 nm) with a maximum power delivery to the sample of 4 mW. A laser clean-up filter was used in the excitation path, combined with a 540 nm dichroic beam splitter and a 532 nm blocking notch filter in the collection optics. When recording a PL image, a 650 nm long-pass filter is inserted in the fluorescence collection path to block the diamond Raman emissions (all filters were supplied by Semrock).

Photon autocorrelation measurements were carried out using the Hanbury-Brown and Twiss method with a 50:50 beam splitter and two single photon detecting diodes with timing resolution ~500 ps. Background signals were measured by recording the fluorescence signal from the bulk diamond at a position misaligned from the laser writing site and were then used to calculate the baseline for the autocorrelation dataset according to the function

$$a = 1 - \left(\frac{S}{S+B}\right)^2$$

where $S$ and $B$ are the photon count rates for the emitter and the background fluorescence respectively. The raw datasets, $g^{(2)}_{raw}(\tau)$, are then corrected for this baseline according to

$$g^{(2)}(\tau) = \frac{g^{(2)}_{raw}(\tau) - a}{1 - a}$$

The datasets, shown in Figure S4, are then fit with a function for the photon autocorrelation of a three-level system, which is established using a system of coupled rate equations

$$g^{(2)}(\tau) = 1 - ce^{-|\tau|/\tau_2} + (c-1)e^{-|\tau|/\tau_3}$$

where c, $\tau_3$ and $\tau_2$ are constants related to the internal dynamics of the NV center.

Polarization measurements of the NV fluorescence are taken by placing a linear polarizing filter (Thorlabs LPVIS100) before a single photon detecting diode. The fluorescence intensity is measured as a function of polarizer angle, $\theta$, using a motorized rotation mount (Thorlabs K10CR1/M). The measurements for each center in the array are presented in figure S3. The data are fitted according to a Malus-law function

$$I(\theta) = I_0 + \Delta I \cos^2(\theta + \varphi)$$

where $I_0$ is the minimal intensity, $(I_0 + \Delta I)$ is the peak intensity and $\varphi$ represents orientation of the of emission dipole projection on the plane of the microscope.

The position in the image plane of each NV center was determined from the PL images by fitting 2D Gaussian surfaces to the measured intensity distributions. The spatial resolution of the microscope is approximately 200 nm (as measured by a Gaussian fit to the image of a single NV⁻ center). However, the knowledge that we are investigating single emitters enables localization with precision much higher than the intrinsic resolution of the microscope. In our case, repeatedly fitting three emitters in a diamond (using one as a "reference" and measuring the distance between the reference and the target emitters each time) yielded a repeatability of approximately 20 nm in *x,y* for timescales consistent with the localization measurements (approximately 20 minutes). The target grid was determined by performing a least-squares fit of a regular rectangular grid to the PL image using fitting parameters of offset in *x* and *y*, spacings in *x* and *y*, shear and rotation.

Figure S2 shows the pulse sequence used for Hahn Echo measurements of the spin coherence time. A biasing dc magnetic field was applied parallel to the axis of the NV center to split the $m_s = \pm 1$ levels in the electronic ground state. The alignment of this bias field was achieved using the magneto-optic effect: an approximate <111> alignment was determined by continuous wave optically detected magnetic resonance at low field (~5 mT), and then the field strength was increased to approximately 40 mT. The magnet was manipulated using powered three-axis stages until the emitter count rate did not drop between 5 and 40 mT: this corresponds to a field alignment along the axis within approximately 1 degree. A Laser Quantum Gem 532 (100 mW) was used through an Isomet 1250C acousto-optic modulator operated with a custom driver to eliminate the leakage of green laser light. The sample was mounted onto a co-planar waveguide board, and microwaves were delivered to the sample using a 20 µm copper wire over the surface. A Keysight N5172B source was switched using a Swabian Instruments PulseStreamer and subsequently amplified with a MiniCircuits ZHL-16W-43-S+ amplifier. A high-NA (1.4) oil immersion objective was used for both excitation and collection paths. The collected light was focused through a pinhole onto a pair of Excelitas NIR-enhanced single photon detectors and recorded with a Swabian Instruments Timetagger. The spin echo signal was fitted to a stretched exponential decay function

$$f(t) = e^{-\left(t/T_2\right)^\alpha}$$

where $t$ is the total free-precession time, indicated in figure S2, $T_2$ is the electron spin-coherence time and $\alpha$ is a free-parameter related to the dynamics of the spin environment [1].

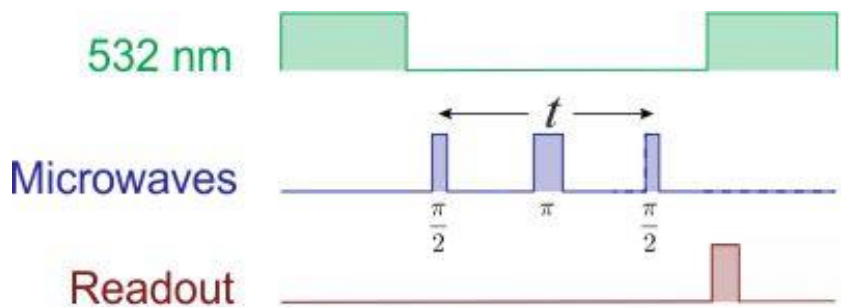

Figure S2: Pulse sequence for Hahn echo measurements used to record spin coherence as shown in Figure 1D.

**Supplementary data:**

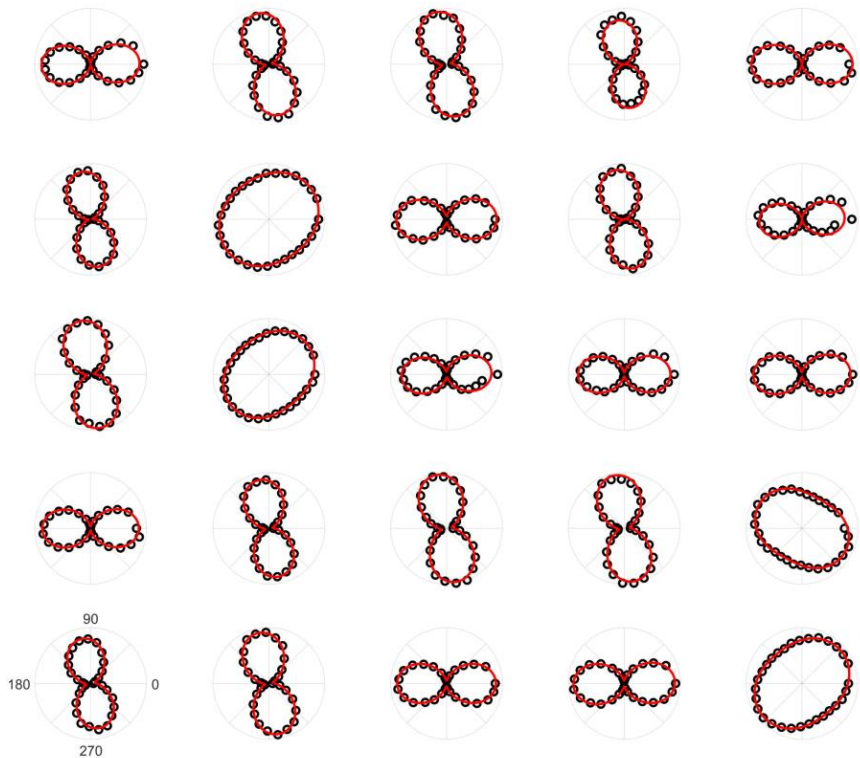

Figure S3: Polar plots of fluorescence intensity revealing the orientation of the 25 NV centers as viewed in the array in figure 1B. Highly polarized fluorescence indicates an NV centers lying in the plane of the microscope image while those with less polarized fluorescence lie out of the plane (see Fig 4C, D and E). The site at the right-hand end of the row second from bottom contains two NV centres, one of each in-plane orientation.

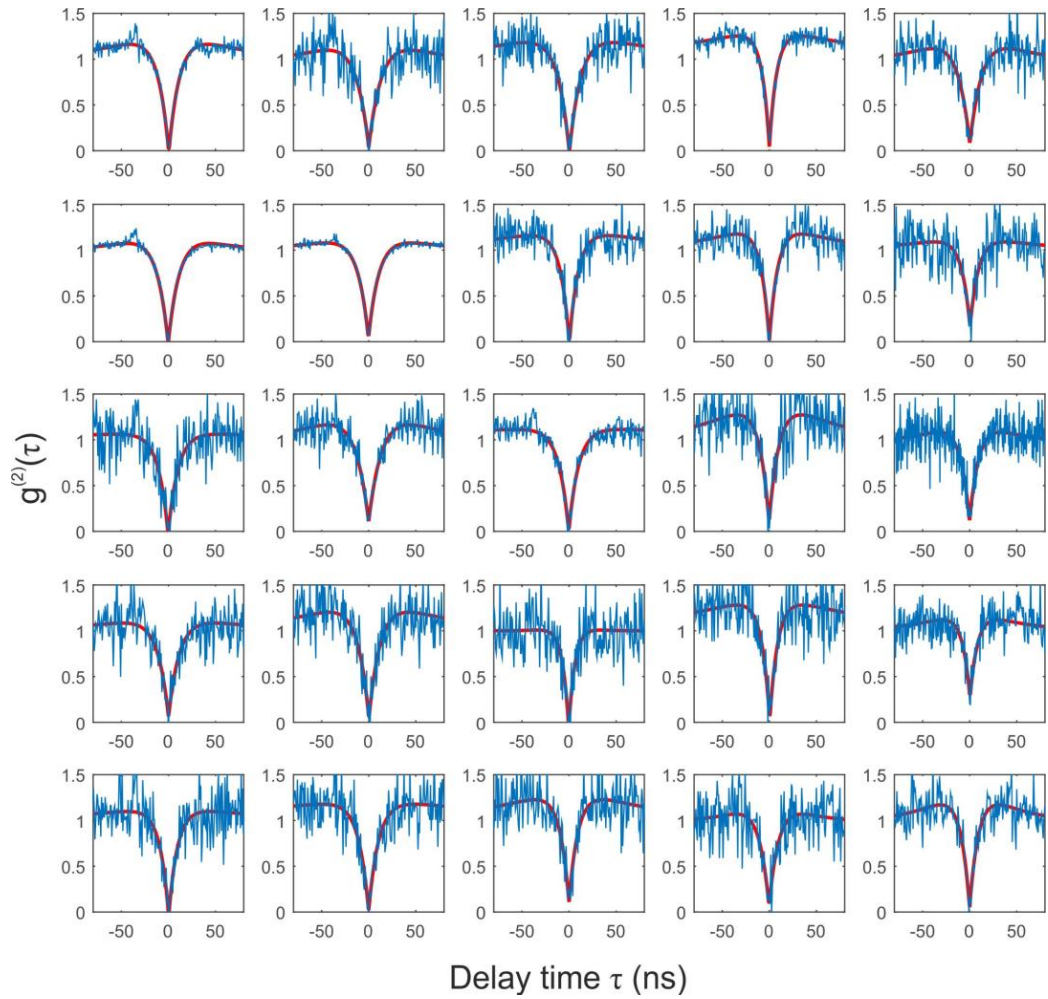

Figure S4: Background-corrected photon correlation histograms for each of the laser-written sites as viewed in the array in Fig 1B (see Methods for correction method).

**References and Notes:**

24. [1] N. Bar-Gill *et al.*, *Nature Communications*. **4**, 1743 (2013).